\documentclass[useAMS, usenatbib, usegraphicx]{mn2e}

\usepackage{amssymb, amsmath}
\usepackage{bm}

\title[Fermi Bubbles by Leptonic AGN Jets]
{The Fermi Bubbles: Gamma-ray, Microwave, and Polarization Signatures of Leptonic AGN Jets}

\author[H.-Y.\ K.\ Yang et al.]
{H.-Y.\ Karen Yang,$^1$\thanks{Email: hsyang@umich.edu},
M.\ Ruszkowski $^{1,2}$,
and E.\ Zweibel $^3$ \vspace{8pt} \\
$^1$Department of Astronomy, University of Michigan, Ann Arbor, MI\vspace{5pt}\\
$^2$The Michigan Center for Theoretical Physics, Ann Arbor, MI\vspace{5pt}\\ 
$^3$Department of Astronomy and Physics and Center for Magnetic Self-Organization, University of Wisconsin-Madison, Madison, WI}

\date{2013 February}

\pagerange{\pageref{firstpage}--\pageref{lastpage}} \pubyear{2013}

\def\LaTeX{L\kern-.36em\raise.3ex\hbox{a}\kern-.15em
    T\kern-.1667em\lower.7ex\hbox{E}\kern-.125emX}

\begin{document}

\label{firstpage}

\maketitle

\begin{abstract}

The origin of the two large bubbles at the Galactic center observed by the {\it Fermi Gamma-ray Space Telescope} and the spatially-correlated microwave haze emission are yet to be determined. To disentangle different models requires detailed comparisons between theoretical predictions and multi-wavelength observations. Our previous simulations, which self-consistently include interactions between cosmic rays (CRs) and magnetic fields, have demonstrated that the primary features of the {\it Fermi} bubbles could be successfully reproduced by recent jet activity from the central active galactic nucleus (AGN). In this work, we generate gamma-ray and microwave maps and spectra based on the simulated properties of cosmic rays and magnetic fields in order to examine whether the observed bubble and haze emission could be explained by leptons contained in the AGN jets. We also investigate the model predictions of the polarization properties of the {\it Fermi} bubbles, including the polarization fractions and the rotation measures (RMs). We find that: (1) The same population of leptons can simultaneously explain the bubble and haze emission given that the magnetic fields within the bubbles are very close to the exponentially distributed ambient field, which can be explained by mixing in of the ambient field followed by turbulent field amplification; (2) The centrally peaked microwave profile suggests CR replenishment, which is consistent with the presence of a more recent second jet event; (3) The bubble interior exhibits a high degree of polarization because of ordered radial magnetic field lines stretched by elongated vortices behind the shocks; highly-polarized signals could also be observed inside the draping layer; (4) Enhancement of RMs could exist within the shock-compressed layer because of increased gas density and more amplified and ordered magnetic fields, though details depend on projections and the actual field geometry. We discuss the possibility that the deficient haze emission at $b<-35^\circ$ is due to the suppression of magnetic fields, which is consistent with the existence of lower-energy CRs causing the polarized emission at 2.3\ GHz. Possible AGN jet composition in the leptonic scenario is also discussed. 

\end{abstract}



\section{Introduction}  


One of the most important discoveries of the {\it Fermi Gamma-ray Space Telescope} is the two giant bubbles that are nearly symmetric about the Galactic center (GC), extending to $\sim 50^\circ$ above and below the Galactic plane \citep{Su10}. The gamma-ray emission of the {\it Fermi} bubbles is observed in the energy range of $1\leq E_\gamma \leq 100$\ GeV and has an almost spatially uniform hard spectrum, sharp edges, and an approximately flat brightness distribution. The bubbles are spatially coincident with the `microwave haze', which was identified by the {\it Wilkinson Microwave Anisotropy Probe} ({\it WMAP}; \cite{Finkbeiner04}) and recently confirmed by the {\it Planck} satellite \citep{PlanckHaze}. In contrast to the flat brightness profile of the gamma-ray bubbles, the brightness distribution of the haze decreases with increasing $|b|$ \citep{Dobler08}, where $b$ is the Galactic latitude, and exhibits a sharp cutoff for $b<-35^\circ$ \citep{Dobler12a}. Recently, \cite{Hooper13} compiled the latest data from the 4.5-year {\it Fermi} observations and confirmed the uniformly hard spectrum of the bubbles after removing possible excessive gamma-ray emission from dark matter annihilation close to the GC.    
Though the 7-year {\it WMAP} measurements do not show evidence of polarization for the haze \citep[e.g.,][]{Gold11}, possibly hidden by the noise \citep{Dobler12a}, recently \cite{Carretti13},
observing with {\it S-PASS}, discovered a high degree of polarized lobe emission  at 2.3\ GHz, with similar morphology as the {\it Fermi} bubbles but extending to $|b|\sim 60^\circ$ and also to the side of the bubbles. Finally, the rims of the bubbles are also correlated with arc features in the {\it ROSAT} X-ray map at 1.5\ keV \citep{Snowden97, Su10}. These spatially resolved multi-waveband observations together provide valuable information about the physical origin of the {\it Fermi} bubbles.  


The unique morphology and symmetry of the {\it Fermi} bubbles about the Galactic plane suggest that they originate from some episode of energy injection at the GC, possibly associated with nuclear star formation processes \citep{Crocker11, Carretti13}, or with past activity of the central AGN \citep[][hereafter Y12]{Cheng11, Zubovas12, Guo12a, Yang12}. The gamma-ray emission can be produced by the decay of neutral pions following inelastic collisions between cosmic ray protons (CRp) and thermal nuclei (i.e., the `hadronic' model), and/or inverse Compton (IC) scattering of photons in the interstellar radiation field (ISRF) and the cosmic microwave background (CMB) by cosmic ray electrons (CRe) (i.e., the `leptonic' model). For the latter, various mechanisms for the source of CRe have been proposed, including cosmic ray (CR) acceleration within AGN jets (\cite{Guo12a}; Y12), shocks \citep{Cheng11}, or plasma wave turbulence \citep{Mertsch11}. In order to disentangle these processes, it is crucial for each model to make predictions for the {\it spatial} and {\it spectral} features of the bubble emission that could be tested against the multi-wavelength observational data.      


In our previous work (Y12), we showed that the key characteristics of the {\it Fermi} bubbles and the {\it ROSAT} X-ray features can be successfully reproduced by recent AGN jet activity. Using three-dimensional (3D) magnetohydrodynamic (MHD) simulations that include the dynamical coupling between CRs and thermal gas, and anisotropic CR diffusion along the magnetic field lines, we self-consistently accounted for the smooth surface, sharp edges, the projected size and shape of the bubbles, and the projected location of the shocks. \footnote{We note that in the AGN jet model (\cite{Guo12a}; Y12), the location of shocks matches with the {\it outer ROSAT} X-ray arc feature (Figure 20 in \cite{Su10}) and the contact discontinuities are marked by the observed bubble edges. In contrast, models that are based on in-situ CR acceleration \citep{Cheng11, Mertsch11} assume that the shock fronts coincide with the bubble edges. Future multi-wavelength observations will help determine the actual location of the shock fronts and distinguish different models.} The broad agreement between our model predictions and the observed bubble features provides  supporting evidence for the AGN jet scenario and demonstrates the importance of self-consistent modeling of the interaction between CRs and magnetic fields. In this work we make further contact with observations by exploring our model predictions for the emission properties of the bubbles in multiple wavebands. 


The predicted emission properties depend on whether the jets are leptonic or hadronic. Since the composition of AGN jets is still largely unknown, one may make assumptions about the composition and compare the models based on these assumptions to the data. Assuming the gamma-ray and microwave emitting CRs are leptons, \cite{Su10} and \cite{Dobler12a} showed that the bubble and haze emission can be successfully explained by the same CR population. 
Using 2D non-MHD simulations, \cite{Guo12a} also showed that the leptonic jet model is in agreement with the gamma-ray spectrum and the observed microwave emission at 23\ GHz, though the hadronic model may explain the gamma-ray emission too. We note that the previous analyses were usually done by taking a single value for the ISRF or magnetic field at a given Galactic latitude integrated over an arbitrary path length. However, as demonstrated in Y12, since the size of the {\it Fermi} bubbles is comparable to the distance from the Sun to the GC, it is crucial to properly account for the effect of 3D projection. 


Using simulated 3D distributions of CRs and magnetic fields, in this work we explore the predictions of the leptonic jet model and identify the critical mechanisms for explaining the {\it morphological, spectral, and polarization} properties of the {\it Fermi} bubbles and microwave haze. Since the gamma-ray and microwave radiation predicted by the hadronic model would require detailed modeling of the secondary electrons and positrons generated during the decay of charged pions, we defer the investigation of the hadronic model to future work. 

The structure of the paper is as follows. In \S~\ref{sec:method}, we describe the details of our MHD simulations and the method to generate simulated gamma-ray and microwave maps and spectra. In \S~\ref{sec:Bfield}, we present the predicted gamma-ray and microwave spectra in the leptonic scenario and discuss implications for the magnetic field within the bubbles. In \S~\ref{sec:2ndcr}, we compare the simulated gamma-ray and microwave maps with observations and show that CR replenishment is required in order to match the centrally peaked profile of the observed microwave haze. In \S~\ref{sec:polar} we show the model predictions of the polarization properties of the {\it Fermi} bubbles, including simulated maps of polarization fractions and rotation measures. In \S~\ref{sec:discussion}, we discuss magnetic field suppression as the possible cause of the deficient haze emission at $b<-35^\circ$. We then discuss the relative importance of the hadronic components and inferred composition of the AGN jets. Our conclusions are in \S~\ref{sec:conclusion}.  


\section{Methodology}
\label{sec:method}


We perform 3D MHD simulations of CR injection from the GC with self-consistently modeled CR advection, anisotropic CR diffusion along magnetic field lines, and dynamical coupling between the CRs and the thermal gas. The CRs are modeled as a second fluid in the MHD equations. In this approach, the dynamical effect of CRs is treated by following the evolution of the CR energy density, with no assumptions about the spectrum or species of the CRs. The CR diffusion coefficient parallel to the magnetic field lines is set to the typical value in the Galaxy, i.e., $\kappa_\parallel = 4\times 10^{28}\ {\rm cm}^2\ {\rm s}^{-1}$, whereas the perpendicular diffusion is assumed to be negligible. Our CR treatment neglects CR acceleration and streaming. These processes will be included in future work. We use the adaptive-mesh-refinement FLASH v.4 code \citep{Flash, Dubey08, Lee09, Lee13}. We refer the reader to Y12 for the detailed code description, the initial conditions, and the numerical techniques of the simulations. Here we only emphasize the parts of the setup and parameters that are different from the previous simulations.     


In Y12, the initial magnetic field\footnote{The magnetic field in our simulations refers to the large-scale regular or mean field component of the Galactic magnetic field, rather than the small-scale turbulent field.} is assumed to be tangled at a given coherence length $l_{\rm B}$ with a constant average field strength. Such a configuration has allowed us to qualitatively study the effect of magnetic draping during the supersonic bubble expansion and explain the sharp edges of the observed bubbles. However, in order to quantify the synchrotron radiation as a function of position, it is essential to start the simulation with a more realistic magnetic field distribution. To this end, for our initial magnetic field we adopt the default exponential model in GALPROP \citep{Strong07} which has the following spatial dependence,
\begin{equation}
|B| = B_0 \exp (-z/z_0) \exp (-R/R_0),
\label{eq:Bfield}
\end{equation} 
where $R=\sqrt{x^2+y^2}$ is the projected radius to the Galaxy's rotational axis, $B_0$ is the average field strength at the GC, and $z_0$ and $R_0$ are the characteristic scales in the vertical and radial directions, respectively. We adopt $z_0=2$\ kpc and $R_0=10$\ kpc, which are best-fit values in the GALPROP model to reproduce the observed large-scale 408\ MHz synchrotron radiation in the Galaxy. We choose $B_0=50\ \mu$G based on the observed field strength at the GC \citep{Crocker10}. 

The assumption of a constant coherence length for the Galactic magnetic field may be somewhat unrealistic. While the magnetic field in the Galactic halo may be coherent on large scales and possesses, e.g., a dipole or quadrupole structure, the disk field is likely to be more tangled on small scales with a few field reversals \citep[see reviews by][]{Brown10, Noutsos12}. We mimic such dependence by superposing a `disk' field, which has a coherence length of $l_{\rm B,disk}$ with spatial dependence defined in Eq.\ \ref{eq:Bfield}, and a `halo' field, which has $l_{\rm B,halo}$ with a constant average field strength of $1\ \mu$G. The magnetic field initialized in this manner is dominated by the disk field near the Galactic plane, whereas the halo field is stronger at higher $|b|$. 

For a given magnetic field coherence length $l_{\rm B}$, the tangled field is initialized in the same way as in Y12, but at the last step after performing the inverse Fourier transform, the field strength is normalized according to the desired spatial dependence. The procedure of divergence cleaning and normalization is then performed iteratively until $\nabla \cdot {\bf B}$ vanishes. The disk field is initialized following the above method, while the halo field has vanishing divergence by construction. We note that the superposed field remains divergence free.  

The injection of CRs from the GC is performed using the same implementation and jet parameters as in Y12. These parameters are carefully chosen in order to match the observed morphology of the {\it Fermi} bubbles at the end of the simulation ($t=1.2$\ Myr), the limb-brightened intensity distribution in the {\it ROSAT} X-ray 1.5\ keV map, and the gas temperature inside the bubbles inferred from X-ray line ratios \citep{Miller13}. As discussed in detail in Y12, these criteria give very stringent limits on the jet parameters; varying any of them would easily violate one of the observational constraints. \footnote{For instance, when a faster jet speed is used \citep[e.g.,][]{Guo12a}, the gas temperature inside the bubbles would be too hot to be consistent with the enhanced OVIII to OVII line ratios for sight lines passing through the bubbles. Note that the jet speed has a critical impact on the simulated CR distribution, age, and hydrodynamic properties of the bubbles (see Y12 for detailed discussion).} The largest uncertainty is in the assumed initial hot gas density profile. Since the estimated jet power and total pressure contained within the bubbles are directly proportional to the gas densities near the core, they are likely overestimated because of the cuspiness of the initial gas density profile.  

As we will discuss in \S~\ref{sec:2ndcr}, the observed synchrotron profile suggests the existence of a second CR population, possibly related to the more recently discovered gamma-ray jets found by \cite{Su12}. Therefore, for one of the runs we inject a second jet at $t=0.7$\ Myr with $1/3$ of the jet speed and 2/5 of the jet radius with respect to the first jet (the other parameters stay the same as for the first jet). The power of the second jet is thus only 1/7.5 of the first jet. These parameters are chosen in order to match the observed synchrotron emission profile. Note however that the parameters of the second jet are not expected to be unique due to degeneracies among jet parameters when morphology is the only constraint (\cite{Guo12a}; Y12). The magnetic field and jet parameters for the simulations presented in this paper are summarized in Table \ref{tbl:params}.

\begin{table}
\caption{Simulation Parameters}
\begin{center}
\begin{tabular}{ccccc}
\hline
Run & Magnetic Field & $l_{\rm B,disk}$ (kpc) \footnotemark[1] & $l_{\rm B,halo}$ (kpc) \footnotemark[2] & \# Jets \\ 
\hline
\hspace{5pt} A \footnotemark[3] & Single & - & 9 & 1 \\
B & Single & 9 & - & 1 \\
C & Superposed & 1 & 9 & 1 \\
D & Superposed & 0.5 & 9 & 1 \\
E & Superposed & 0.5 & 9 & 2 \\
\hline
\multicolumn{5}{l}{\footnotesize \footnotemark[1] Magnetic field having a spatial dependence defined in Eq.\ \ref{eq:Bfield}.}\\
\multicolumn{5}{l}{\footnotesize \footnotemark[2] Magnetic field having a constant average strength.}\\
\multicolumn{5}{l}{\footnotesize \footnotemark[3] Same as Model D in Y12.}
\end{tabular}
\end{center}
\label{tbl:params}
\end{table}


The simulated gamma-ray and microwave maps and spectra are computed by post-processing of the 3D distribution of CRs at the end of the simulations. This is a good first-order approximation because the dynamical time of bubble expansion is always shorter than the IC and synchrotron cooling time estimated from our simulations, except for the very early stage of the bubble evolution ($t \lesssim 0.1$\ Myr for 10\ TeV CRe; see \S~\ref{sec:conclusion} for more discussion). In the leptonic scenario, the gamma-ray emission comes from IC scattering of the ISRF and CMB photons by CRe, and the microwave emission originates from synchrotron radiation of CRe gyrating along magnetic field lines. Since our MHD simulations do not model the spectrum of CRs directly, we follow \cite{Su10} and assume that the CR spectrum follows a power-law distribution with a spectral index of $-2$ and ranges from $0.1-1000$\ GeV. The gamma-ray emissivity is computed for each computational volume element in our simulations using the Klein-Nishina IC cross section \citep{Jones68} based on the simulated CR number density and the ISRF model adopted from GALPROP v.50 \citep{Strong07}. The gamma-ray intensity maps and spectra are then generated by projecting the gamma-ray emissivities along lines of sight across the sky. Similarly, using standard formula for the synchrotron emission \citep[e.g.,][]{Ginzburg79, Strong00}, the synchrotron maps and spectra are obtained according to the CR distribution and energy densities of the magnetic field at the end of the simulations.   


\section{Results}
\label{sec:result}

\subsection{Bubble and haze spectra -- implications for magnetic fields within the bubbles}
\label{sec:Bfield}

In this section, we compare the simulated gamma-ray and microwave spectra based on the leptonic AGN jet model with the observed spectra of the {\it Fermi} bubbles and the microwave haze. We demonstrate that in order for the same CR population to simultaneously reproduce both the bubble and haze emission, the magnetic field inside the bubbles has to be very close to the initial ambient values, which can be explained by mixing in of the ambient field followed by turbulent field amplification. 

Our previous simulations in Y12 have reproduced a broad range of properties of the observed {\it Fermi} bubbles, including their projected size and shape, smooth surface, and sharp edges. Therefore, it is instructive to compute the gamma-ray and microwave emission based on the 3D CR distribution in the simulations. However, since these simulations did not start with a realistic distribution of magnetic field but assumed constant average field strength and coherence length, we first obtain preliminary microwave spectra using the exponential model (Eq.\ \ref{eq:Bfield}) as an approximation for the magnetic field inside the bubbles at the end of the evolution. The CR distribution is adopted from Run A in Table \ref{tbl:params} (same as Model D in Y12). 

\begin{figure}
\begin{center}
\includegraphics[scale=0.85]{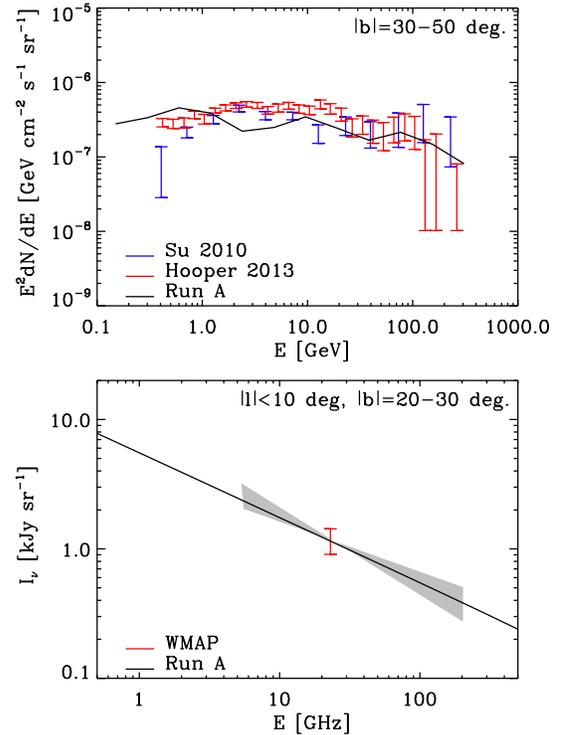} 
\caption{{\it Top}: Simulated gamma-ray spectra at $t=1.2$\ Myr for emission integrated over $|b|>30^\circ$, overplotted with the observed data of the {\it Fermi} bubbles \citep{Su10, Hooper13}. {\it Bottom}: Microwave spectra averaged over $|l|<10^\circ$, $20^\circ<|b|<30^\circ$. The data point represents the {\it WMAP} data in the 23\ GHz $K$ band and the shaded area indicates the range of synchrotron spectral indices allowed for the {\it WMAP} haze \citep{Dobler08}.}
\label{fig:spectra}
\end{center}
\end{figure}

Figure \ref{fig:spectra} shows the simulated gamma-ray (top) and microwave (bottom) spectra averaged over the same patch of the sky as used in previous observational studies. As previously found \citep{Su10, Dobler12a}, a CR spectrum of slope $-2$ provides a good match to the observed hard spectrum of the bubbles and haze. \footnote{Recently, \cite{Hooper13} analyzed the bubble spectra as a function of Galactic latitudes and found a best-fit slope of $-3$ for the CR spectrum. However, the latitude dependence is sensitive to the modeling of the excessive gamma-ray signal close to the GC, and also to the uncertainties in the subtraction of various components near the Galactic disk. Therefore, in this study we focus on the comparison with the latitude-integrated bubble and haze spectra.} 
By comparing the amplitudes of the simulated and observed gamma-ray spectra, we find that only a small fraction, $f_{\rm e,\gamma}=4.0\times 10^{-4}$, of the total (regardless of species) CR energy density in our simulation, $e_{\rm cr,sim}$, is needed in order for the model to match the observed data\footnote{Note that in our simulations, the total CR energy density is degenerate with the thermal energy density (Y12). Therefore, $f_{\rm e,\gamma}$ serves only as a convenient parameter for measuring the required amount of CRe to match the observed emission, rather than the actual fraction of CRe in the (unconstrained) total CR population.}, which is consistent with the result of \cite{Guo12a}. Similarly, for the microwave spectra we find that only a fraction of $f_{\rm e,\nu}=6.0\times 10^{-4}$ of the total CR energy density needs to be provided by CRe in order for the simulation to reproduce the observed haze emission. We use different normalization factors for the gamma-ray and microwave emission in order to allow for uncertainties in the actual magnetic field strength, and for differences due to projections of our symmetric CR and magnetic field distributions as opposed to the asymmetric {\it Fermi} bubbles that bent slightly to the west. Despite the uncertainties, for the leptonic model to be considered successful, the two normalization factors, $f_{\rm e,\gamma}$ and $f_{\rm e,\nu}$, are not expected to differ by more than a factor of a few.     

These similar values of $f_{\rm e,\gamma}$ and $f_{\rm e,\nu}$ have two important implications. First, they imply that the emission of the {\it Fermi} bubbles and the microwave haze can be produced by the same leptonic CRs, as previously suggested \citep{Su10, Dobler12a}. However, we note that results of the previous observational studies were based on assumed values for the ISRF and magnetic field integrated over an arbitrarily chosen path length, whereas our simulated spectra are computed taking into account line-of-sight projections of the 3D distributions of the magnetic field and self-consistently simulated CRs through the simulated size of bubbles. The good agreement between the simulated and observed spectra provides support for the 3D CR distribution and bubble size derived from our simulations.

Secondly, this simple exercise of matching the amplitudes of spectra implies that the magnetic field as described by the exponential model is approximately a {\it lower} limit for the magnetic field {\it in the bubble interior}. If the bubble field were much smaller than the model value, more CRe would be needed to match the observed microwave emission, which would however overproduce the IC radiation in the gamma-ray waveband. On the other hand, if the magnetic field inside the bubbles is somehow much greater than the model value (though this is rather unlikely, if no additional mechanisms are invoked to generate magnetic fields over the course of bubble expansion, as will be discussed later), less CRe would be required for the haze emission, and in this case the observed gamma-ray bubbles would also need partial contribution from other physical sources, such as hadronic processes. Therefore, in the {\it purely} leptonic scenario, the magnetic field strength has to be very close to the exponential model values. 

Such a large magnetic field strength inside the bubbles is somewhat counter-intuitive, because effects such as magnetic draping \citep{Lyutikov06, Ruszkowski07, Ruszkowski08, Dursi08} and adiabatic expansion would act to reduce the magnetic energy density within the bubbles. In our simulations, the magnetic energy injected with the AGN jets is constrained by the observed magnetic field strength at the GC and is only a small fraction ($\sim 10^{-3}$) of the total jet power (Y12). The injected field is diluted by the adiabatic expansion and thus has little contribution to the magnetic field inside the bubbles. One test run without magnetic field injection shows almost identical field strength and geometry as the fiducial run, only except for a small region very close to the GC. Therefore, the magnetic field within the bubbles is primarily determined by the response of the ambient field to the bubble expansion.      

\begin{figure}
\begin{center}
\includegraphics[scale=0.9]{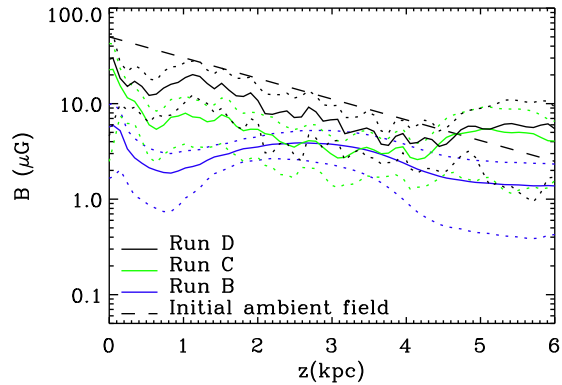} 
\caption{Simulated magnetic field strength at $t=1.2$\ Myr as a function of vertical height from the GC for different coherence length $l_{\rm B}$ (see Table \ref{tbl:params}). The solid and dotted lines show the average and standard deviation, respectively, of the field strengths inside the region $|x|<0.5$\ kpc and $|y|<0.5$\ kpc. Magnetic fields with larger $l_{\rm B}$ (e.g., Run B) result in smaller field strength within the bubbles due to longer field amplification timescales, whereas fields with smaller $l_{\rm B}$ (e.g., Run D) are more efficiently amplified to values comparable to the ambient field (see text for details).} 
\label{fig:Bprf}
\end{center}
\end{figure}

In order to investigate the evolution of the magnetic field within the bubbles, we performed simulations with varying initial coherence length (Run B, C, and D in Table \ref{tbl:params}). Figure \ref{fig:Bprf} shows the magnetic field strength as a function of vertical height from the GC (the height of the bubbles is $\sim 6$\ kpc). When the coherence length $l_{\rm B}$ is large (e.g., Run B), the bubble field is weaker than the ambient field, whereas for smaller $l_{\rm B}$ (e.g., Run D), the field strength can be comparable to the ambient value (see also Figure \ref{fig:Bmap}).

One possible cause for the difference in the bubble field for different $l_{\rm B}$ is the level of mixing. During the supersonic bubble expansion driven by the AGN jets, the magnetic field is compressed into shells at the edge of the bubbles. Due to the magnetic draping effect, the magnetic field becomes aligned with the bubble surface and the field strength is amplified. The effect of draping is more efficient when the direction of the initial field is parallel to the bubble surface. Therefore, for the tangled magnetic fields in our simulations, the draping effect is generally more pronounced if the coherence length $l_{\rm B}$ is larger, whereas for smaller $l_{\rm B}$ the fields are more randomly oriented on small scales and the components perpendicular to the bubble surface drape less efficiently. Efficient magnetic draping could stabilize small-scale Kelvin-Helmholtz (KH) instabilities \footnote{As shown in Y12, large-scale KH instabilities do not have sufficient time to develop because the bubbles are young; however, the instabilities do occur on small scales as the KH timescale is proportional to the wavelength of the perturbations (Eq.\ 19 in Y12). Magnetic field does not suppress KH instability for perturbations perpendicular to the magnetic field. We only consider the KH instabilities because the estimated Rayleigh-Taylor timescale is much longer than the age of the bubbles.} and inhibited mixing between the bubbles and the ambient medium \citep{Lyutikov06, Ruszkowski07}, resulting in weaker field strength within the bubbles. However, using passively evolving tracer particles for the ambient medium, we found that the level of mixing is similar for different $l_{\rm B}$. Specifically, for all the runs the fraction of ambient gas is $\gtrsim 10\%$ near the bubble edges and gradually decreases toward the bubble center. The runs with smaller $l_{\rm B}$ indeed have more mixing, but the differences are only a few percent, which is insufficient to explain the difference in the magnetic field strength within the bubbles. This implies that some other mechanism amplifies the mixed-in field to the ambient value in the runs with small $l_{\rm B}$. 
We note that the mixing process could potentially bring the interstellar CRp into the bubbles and produce gamma-ray emission via the hadronic process. We estimated the expected emission from the mixed-in CRp and found their contribution is small compared to the bubble emission (see Appendix \ref{appendix2}). 

Amplification of magnetic fields can occur when the field lines are stretched and wound up by vortical motions of the gas \citep[e.g.,][]{Zweibel03}. Although the ambient medium does not have an explicitly driven turbulence at the beginning of the simulations, the tangled magnetic fields provide tension that stirs up the gas. After equipartition is established, the spectrum of velocity fluctuations is similar to that of the magnetic field and peaks at the scale of the initial magnetic field coherence length (see Eq.\ 12 and 13 in Y12). As a result, the RMS value of the random velocity field for the ambient medium ($\sigma_{\rm amb}$) measured at a fixed scale smaller than $l_{\rm B}$ is greater for fields with smaller initial coherence length $l_{\rm B}$. In particular, $\sigma_{\rm amb}$ for Run B is only a few $\mbox{km\ s}^{-1}$, while $\sigma_{\rm amb}$ for Run C and D is on the order of tens of $\mbox{km\ s}^{-1}$ measured on a scale of 0.1\ kpc. The latter is comparable to velocity dispersion measurements of the turbulent interstellar medium (ISM) in the thick disk \citep[the dominant component for $z \lesssim 5$\ kpc;][]{Carollo10}. Run C and D thus allow us to study the role of the turbulent velocity field of the ISM in the process of magnetic field amplification.     

As the shocks pass through the turbulent ambient medium, the RMS value of the ambient velocity field gets amplified behind the shocks to $\sigma_{\rm sh} \sim \sqrt{\alpha} \sigma_{\rm amb}$ ($\alpha>1$), and then gradually decreases with increasing distances from the shock front in the post-shock region \citep{Lee97, Larsson09}. The magnetic field is then tangled and amplified by these vortices. The time scale for field amplification can be characterized by the eddy turnover time, $t_{\rm eddy}=l/\sigma_{\rm sh}$, where $\sigma_{\rm sh}$ is measured on the scale of $l$. For the runs with larger $l_{\rm B}$ (e.g., Run B), $\sigma_{\rm amb}$ is smaller and hence it takes a much longer time for the mixed-in fields to be amplified to the ambient value. On the other hand, magnetic field amplification is more efficient for the runs with smaller $l_{\rm B}$ because of larger $\sigma_{\rm amb}$ (a more quantitive estimate of the amplification timescale is given in Appendix \ref{appendix}). 

When the amplification timescale is short enough compared to the bubble formation time (e.g., Run C and D), the magnetic field can grow until it is in equipartition with the turbulent velocity field. We can thus make a crude estimate for the maximum magnetic energy density within the bubbles based on the following relationships, 
\begin{eqnarray}
e_{\rm B,bub} &\sim& e_{\rm turb,bub} \lesssim e_{\rm turb,sh} = 0.5 \rho_{\rm sh} \sigma_{\rm sh}^2 \nonumber \\  
&\sim& 0.5 (r \rho_{\rm amb})(\alpha \sigma_{\rm amb}^2) = \alpha r e_{\rm turb,amb} \sim \alpha r e_{\rm B,amb}, 
\label{eq:equip}
\end{eqnarray}
where $e_{\rm B}=B^2/8\pi$ is the magnetic energy density, $e_{\rm turb}=0.5 \rho \sigma^2$ is the turbulent kinetic energy density, $\rho$ is the gas density, and the last equality assumes equipartition between the velocity and magnetic fields for the ambient medium. For strong shocks, $r=4$ and $\alpha \sim 2$ \citep[][and references therein]{Larsson09}, and therefore the field strength within the bubbles is expected to be $\lesssim \sqrt{8} \sim 2.8$ of the ambient value. This is consistent with the simulated field strengths for the runs with smaller $l_{\rm B}$ (see Figure \ref{fig:Bprf} and Figure \ref{fig:resol}). Furthermore, this demonstrates that magnetic field amplification is a viable mechanism to explain why the bubble field closely traces the ambient field, which is needed for the leptonic model to simultaneously reproduce the bubble and haze emission. We verified that this conclusion does not depend on numerical resolution (see Appendix \ref{appendix}).

We have established that the field strength inside the bubbles is close to the ambient value as described by the exponential model. Therefore, in the analyses presented in the rest of the paper, unless explicitly mentioned otherwise, we show results for the microwave emission computed using the exponential model. 
We note that by adopting the model field rather than the simulated field, we omit only small-scale fluctuations in the microwave intensity maps, while the overall microwave profiles are similar in the two approaches.

\subsection{Morphology of the bubbles and haze -- implications for CR replenishment}
\label{sec:2ndcr}

In this section, we present the morphological properties of the simulated gamma-ray and microwave maps and compare them with the surface brightness distribution of the {\it Fermi} bubbles and the microwave haze. In particular, we show that the centrally peaked profile of the {\it WMAP} haze requires replenishment of CRs, which may be related to the second pair of jets observed by \cite{Su12}.     

\begin{figure*}
\begin{center}
\includegraphics[scale=0.85]{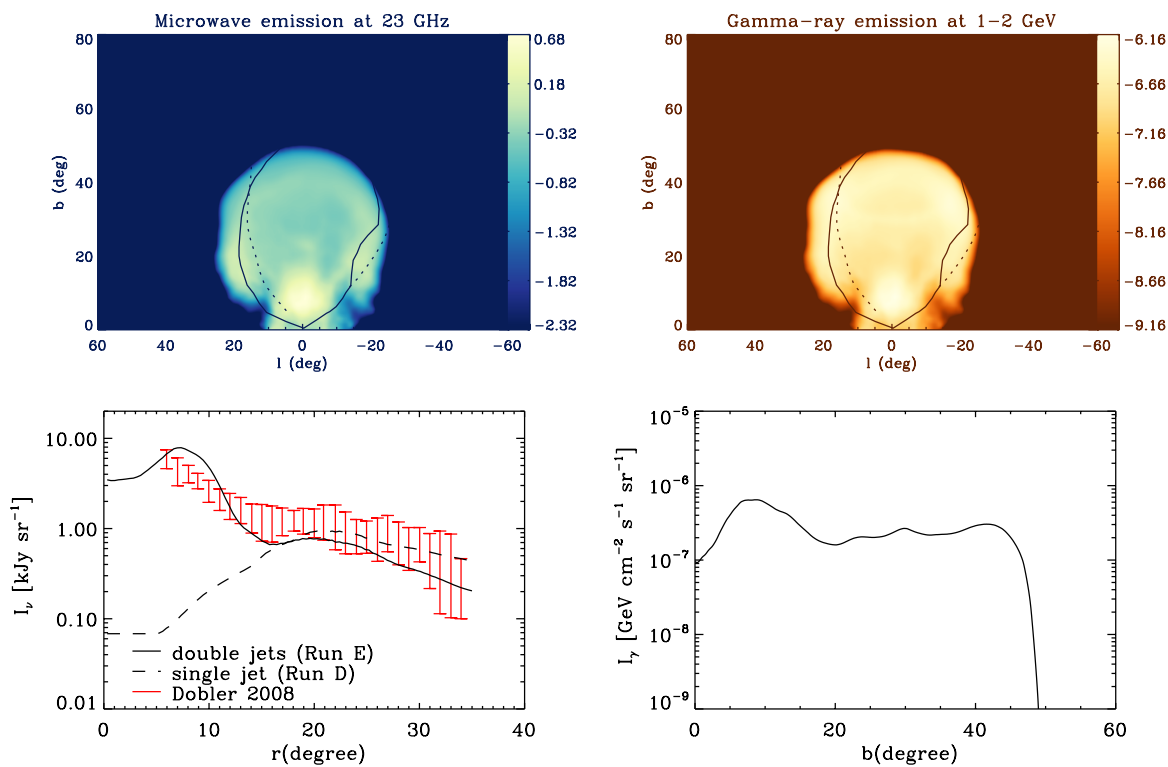} 
\caption{Simulated microwave and gamma-ray intensity maps and profiles for the double-jet simulation (Run E) at $t=1.2$\ Myr. The solid and dotted lines in the upper panels show the surfaces of the observed northern and southern {\it Fermi} bubbles, respectively. The microwave profile (bottom left) is plotted as a function of radial distance from the GC. The centrally peaked profile of the observed microwave haze suggests a replenished CR population at lower $|b|$ (see text). The gamma-ray profile (bottom right) is plotted along the $l=0$ axis and is approximately flat, similar to that observed in the {\it Fermi} bubbles.} 
\label{fig:2jets}
\end{center}
\end{figure*}

While our previous simulation post-processed with the exponential field (Run A in Table \ref{tbl:params}) have provided valuable insights on the required magnetic field distribution within the bubbles, in order to self-consistently incorporate the interaction between the magnetic field and CRs, as an example we now examine the CR distribution using Run D in Table \ref{tbl:params}. This simulation uses the same jet parameters as Run A, but with a more realistic initial magnetic field distribution. These two cases lead to similar general characteristics of the bubble evolution, such as bubble age, shape, sharp edges, and edge-brightened CR distribution. However, because the magnetic pressure inside the bubbles is much greater in Run D than in Run A as a result of mixing and turbulent amplification, the magnetic pressure close to the GC is comparable to the total pressure (plasma $\beta \equiv (P_{\rm th}+P_{\rm cr}+P_{\rm B})/P_{\rm B} \sim 1$, where $P_{\rm th}$, $P_{\rm cr}$, and $P_{\rm B}$ are the thermal, CR, and magnetic pressure, respectively) and contributes to pushing the thermal gas and CRs outward. This causes a deficit of CRs in the innermost $1-2$\ kpc around the GC and hence a depression in the simulated microwave emission profile near the GC (within a radius of $r\lesssim 15^\circ$) shown in Figure \ref{fig:2jets} (dashed line in the bottom left panel). The non-negligible dynamical effect of magnetic fields is robust because the total pressure in our simulations is likely overestimated as it scales with the normalization of the cuspy initial gas density profile (see \S~\ref{sec:method}). If so, the magnetic pressure would be even more dynamically important and cause greater deficit in the microwave emission close to the GC. 

Consequently, the simulated radial profile of the microwave haze is decreasing toward the GC (dashed line in the bottom left panel of Figure \ref{fig:2jets}). Assuming that the bubble field does not go beyond the exponential field (which is used to compute the simulated radial profile), the observed centrally peaked microwave profile implies replenishment of CRs in order to remove the deficit in the microwave emission near the GC. That deficit is caused by the expulsion of CRs by the dynamically important magnetic fields near the GC. If instead of the CR replenishment stronger fields are invoked to compensate for the loss of the microwave emission within $\sim15$ degrees from the GC, then this only exacerbates the problem. That is, the microwave emission actually decreases. Consequently, the above arguments imply that, within central $\sim15$ degrees from the GC, it is CR rather than the magnetic field replenishment that is needed to bring the simulated microwave emission into agreement with observations.

Motivated by this, we included a second, less powerful jet 0.7\ Myr after the first jet injection (Run E, see \S~\ref{sec:method} for details). Figure \ref{fig:2jets} shows the microwave and gamma-ray maps and profiles for this double-jet simulation. With the replenished CR population, the simulated microwave emission at $r \lesssim 15^\circ$ (solid line, bottom left panel) becomes comparable to the observed data of \cite{Dobler08} compiled for the southern {\it WMAP} haze for $-35^\circ<b<0^\circ$, in contrast to the deficit seen in the simulation with a single jet. The observed haze fades away very quickly at latitudes below $b<-35^\circ$ \citep[][Figure 4]{Dobler12a}, which is not seen in our simulated microwave map (top left panel of Figure \ref{fig:2jets}). This suggests that additional physical mechanisms are needed to explain the deficient emission for $b<-35^\circ$, which we discuss in detail in \S~\ref{sec:hibhaze}. 

The simulated gamma-ray surface brightness distribution (top right panel of Figure \ref{fig:2jets}) is approximately uniform, similar to what is seen in the observed {\it Fermi} bubbles. The nearly flat surface brightness distribution results from the edge-brightened underlying CR distribution before line-of-sight projections in our simulations (see Figure 1 of Y12). During the bubble evolution, the CR energy densities injected by the jets generally decrease as the bubbles rapidly expand. Nevertheless, during the active phase of the jets ($t<0.3$\ Myr), the earlier injected decelerating CRs also experience compression by the newly injected material. Consequently, the CRs near the edges and top of the bubbles are more compressed and show higher energy densities than CRs in the central and bottom regions of the bubbles. As a result, the projected CR energy density is roughly flat in the lateral direction and increases at higher $|b|$ (see Figure 2 in Y12). After folding with the decaying ISRF from the Galactic plane, the simulated gamma-ray intensity at $1-2$\ GeV indeed becomes almost uniform also along the $l=0$ axis (bottom right panel of Figure \ref{fig:2jets}). Considering the uncertainties in the initial conditions, as well as possible differences in the results due to 3D projections of our more idealized CR distribution compared to the real distribution in the complex Galactic environment, the agreement between the simulated results and the observations in both the gamma-ray and microwave wavebands is remarkable.  

The inclusion of a second jet event may seem {\it ad hoc} at first glance. However, it is well known that energy injections of supermassive black holes (SMBHs) are episodic with a duty cycle of $\sim 0.1-10$\ Myr \citep{McNamara07}. Therefore, it is possible that Sgr $\mbox{A}^\star$ at the center of our Milky Way has been also going through multiple episodes of activity in the past. In fact, \cite{Su12} recently discovered a pair of gamma-ray jets embedded in the original {\it Fermi} bubbles and tilted at $13^\circ$ from the rotational axis of the Galaxy. The jets are observed at similar energy ranges to the {\it Fermi} bubbles (perhaps extending to lower energies below $\sim 1$\ GeV). The jets have similar gamma-ray intensity to the {\it Fermi} bubbles, in agreement with the intensity of the second jet found in our double-jet simulation (i.e., the bump in the gamma-ray profile at $|b|\lesssim 15^\circ$ in the bottom right panel of Figure \ref{fig:2jets}). If the latest {\it Fermi} data confirms the existence of the jets, they would naturally provide the source of CR replenishment needed to explain the rising microwave profile toward the GC.   


\subsection{Polarization properties}
\label{sec:polar}

The 2.3\ GHz observations by {\it S-PASS} revealed a high degree of polarization in regions largely coincident with the {\it Fermi} bubbles \citep{Carretti13}, suggesting ordered magnetic field geometry inside the bubbles. In this section we use the simulated magnetic fields for Run E to compute the degree of polarization and rotation measures. 

For the calculation of the polarization fractions, we adopt the procedure of \cite{Otmianowska09} and \cite{Skillman13} (see also \cite{Longair94}) to generate the polarization map in the following way. For each simulation cell $i$ along a given line of sight, we integrate the Stokes parameters using
\begin{equation}
\left[ \begin{array}{l} 
I_{i+1} \\ Q_{i+1} \\ U_{i+1} \end{array} \right] = 
\left[ \begin{array}{ccc}
dl & 0 & 0 \\
dl\ f_{p,max} \cos 2\chi & \cos \Delta \phi & -\sin \Delta \phi \\
dl\ f_{p,max} \sin 2\chi & \sin \Delta \phi & \cos \Delta \phi \end{array} \right]
\left[ \begin{array}{l} 
\epsilon_i \\ Q_i \\ U_i \end{array} \right],
\end{equation}
where $\epsilon_i$ is the synchrotron emissivity of cell $i$, $f_{p,max}=(p+1)/(p+7/3)$ is the maximal polarization fraction for CR spectra with power-law index $-p$, $\chi$ is the polarization angle of the electric field (the angle of the projected magnetic field onto the plane of the sky rotated by $\pi/2$), $dl$ is the line-of-sight element for integration, and $\Delta \phi=2.62\times 10^{-17}n_e \lambda^2 B_\parallel$ is the Faraday rotation angle. We assume $p=2$ and $f_{p,max}=0.69$. The polarization fraction is computed from $f_p=\sqrt{Q^2+U^2}/I$, and the polarization angle is $0.5\tan^{-1}(U/Q)$. 

\begin{figure}
\begin{center}
\includegraphics[scale=0.6]{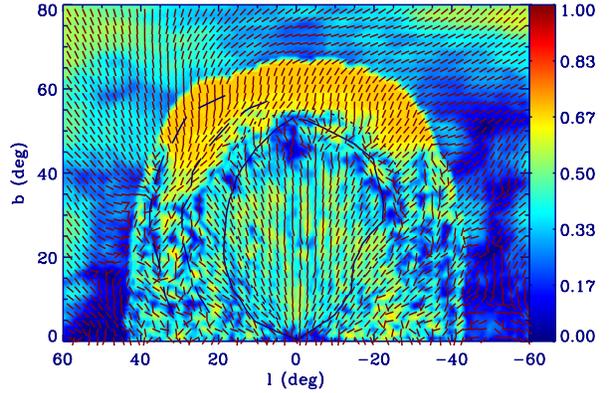} 
\caption{Map of polarization fraction derived from the simulated magnetic fields of Run E at $t=1.2$\ Myr. The vectors overlaid on the map trace the projected magnetic fields ($90^\circ$ degrees from the polarization vector). The solid and dotted lines show the surfaces of the observed northern and southern bubbles, respectively. The long dashed lines show the shock location observed by {\it ROSAT}. The bubble interior reveals high degree of polarization because of the linear structure of magnetic fields amplified by elongated vortices. The shock-compressed region surrounding the bubbles can also have high polarization fractions, especially at higher $|b|$.} 
\label{fig:polar}
\end{center}
\end{figure}

The polarization fractions as a function of positions on the sky are shown in Figure \ref{fig:polar}, overplotted with the vectors tracing the projected magnetic fields (rotated $90^\circ$ from the polarization vector). The results show that the bubble interior indeed can have high degree of polarization, with polarization fractions ranging from $30\ \%$ up to $60\ \%$. This is because the simulated magnetic field lines within the bubbles are quite regular on large scales, preferentially orienting in the radial direction (Y12, Figure 3). 
As the shocks pass through the ambient medium, the RMS velocities not only are enhanced (see discussion in \S~\ref{sec:Bfield}), but also become anisotropic in the post-shock region \citep{Lee97, Larsson09}. The vortices (and thus the magnetic field lines) are stretched in the shock normal direction, allowing more efficient magnetic field amplification along the long axis of the vortices \citep{Schekochihih07}.
This results in linear coherent structures in the magnetic field that are responsible for the large polarization fractions within the bubbles and also within the filamentary structures. These filaments are perhaps related to some of the linear structures (often referred to as {\it Ridges} or {\it Spurs}) in the observed polarization maps \citep{Carretti13}. The observed polarization fractions are $\sim 25\ \%$ averaged over the lobes and $\sim 30\ \%$ for the Ridges, somewhat smaller than the simulated values. The difference is likely due to depolarization by the small-scale turbulent field when projecting along the line of sight \citep[e.g.,][]{Miville08}.     

For the bubble exterior, the synchrotron intensities should be zero because there are no CRs in our models. Nevertheless, in order to still probe the magnetic field geometry, we set the synchrotron emissivity to a small constant value for cells outside the bubble region. As discussed in \S~\ref{sec:Bfield}, since the initial magnetic field is a superposition of the small-scale tangled disk field and the halo field with a larger coherence length, magnetic draping is more effective at higher $|b|$ than close to the Galactic plane. Consequently, the field lines in the draping layer are more ordered at high $|b|$, causing the larger polarization fractions on the top of the bubbles ($|b|\sim 50^\circ-60^\circ$), whereas closer to the Galactic plane, the polarization fraction is patchy and less enhanced (though there is still enhancement compared to the background, i.e., the initial ambient field). Due to magnetic draping, the magnetic field lines tend to align with the bubble surface, and hence the vectors plotted in the figure generally lie parallel to the bubble surface.  However, for some of the regions they do not appear to be perfectly aligned with the projected bubble edges because the projected direction of magnetic fields depends on the actual field orientation within the draping layer. For instance, when the field lines wrap around the bubble surface in the direction perpendicular to the plane of the sky, the polarization vectors for the projected magnetic field look perpendicular to the bubble surface on the map (e.g., $|b|\sim 60^\circ$). 

Note that although our simulations do not predict synchrotron emission outside the bubbles, the above analysis shows that {\it if there exist CRs from other sources}, magnetic draping during the formation of the {\it Fermi} bubbles is able to produce ordered magnetic fields that allow for highly polarized signal between the projected location of the shocks and the bubble edges ($|l|\sim 20^\circ-40^\circ$ in the lateral direction, $|b|\sim 50^\circ-60^\circ$ on the top of the bubbles). This is consistent with the existence of the polarized 2.3\ GHz emission extending up to $|b|\sim 60^\circ$ and also on the sides of the bubbles \citep{Carretti13}.      

\begin{figure}
\begin{center}
\includegraphics[scale=1.0]{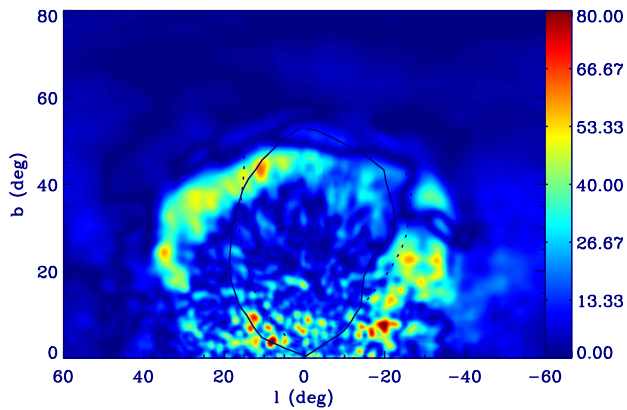} 
\caption{RM map (shown here are absolute values) calculated from gas and magnetic field properties of Run E at $t=1.2$\ Myr. The solid and dotted lines show the surfaces of the observed northern and southern {\it Fermi} bubbles, respectively. The RMs are enhanced because of increased gas density as well as more amplified and ordered magnetic fields within the shock compressed layer, though the actual level of enhancement depends on the exact field geometry within the draping layer.} 
\label{fig:RM}
\end{center}
\end{figure}

Finally, we compute the rotation measures for each line of sight across the sky according to
\begin{equation}
\mbox{RM} = 812 \int n_{\rm e} {\bm B} \cdot \mbox{d} {\bm l} \  \mbox{rad m}^{-2},
\end{equation}
where $n_e$ is the gas electron number density in units of $\mbox{cm}^{-3}$, the magnetic field ${\bm B}$ is in units of $\mu G$, and the line-of-sight element $\mbox{d} {\bm l}$ is in units of kpc. The result is shown in Figure \ref{fig:RM}. For sight lines that solely pass through the background medium, the RM is close to zero because of the randomly oriented ambient magnetic field. For the bubble interior, the RM is also low because the bubbles are very underdense compared to the ambient medium. Parts of the regions surrounding the bubbles, on the other hand, exhibit enhanced values of RM because of increased thermal gas density and magnetic field amplification and alignment due to shock compression and magnetic draping. The actual level of RM enhancement depends on the exact field orientation within the draping layer, i.e., on how much the field lines happen to lie parallel to the lines of sight. Note also that there is RM enhancement near the top of the bubbles, which results from the shock compressed gas cloud on the top of the bubbles (see Figure 1 in Y12) and appears to lie within the boundary of the bubbles due to projection onto the plane of the sky. Thus, although RM enhancement can occur generally between the bubble edges and the shock fronts, the actual areas of enhancement may not lie perfectly along the projected location of the shock fronts or the bubbles edges. These predictions could provide useful information for future RM measurements in the vicinity of the {\it Fermi} bubbles, such as those from the NRAO VLA Sky Survey (NVSS).     

\section{Discussion}
\label{sec:discussion}

\subsection{Deficient haze emission at $b<-35^\circ$}
\label{sec:hibhaze}

 
While the gamma-ray bubbles have a nearly uniform intensity up to $|b|\sim 50^\circ$, the microwave haze above $|b| \gtrsim 35^\circ$ appears to be substantially suppressed; the transition is more clearly seen for the southern haze \citep{Dobler08, PlanckHaze}. Such suppression is not seen in our simulated microwave map (top left panel of Figure \ref{fig:2jets}), suggesting that it originates from some other physical mechanisms which are not included in our simulations. In this section we put together our simulations and the available observational data in order to obtain a consistent picture of the origin of the deficient microwave haze emission. 

Phenomenologically, the deficient haze emission is either caused by the suppression in the CR number density or magnetic fields at high $|b|$. Because the gamma-ray bubbles do not suggest a significant change in CR density above $|b|\gtrsim 35^\circ$, the latter case seems to be more likely \citep{Dobler12a}. Further evidence for this interpretation comes from the recent measurement of polarized emission at 2.3\ GHz by {\it S-PASS} \citep{Carretti13}. The {\it equipartition} magnetic field strength within the polarized lobes is estimated to be $B_{\rm eq}\sim 6\ \mu G$ assuming a path length of 5\ kpc across the size of the lobes and a typical CR proton-to-electron number density ratio of $K=100$. If the system is not in exact equipartition (as is also found to be true for some cluster radio bubbles, e.g., \cite{Dunn04}), given the revised formula for computing the equipartition field strength \citep{Beck05}, the relation between the ratio of the total particle energy ($e_{\rm cr}$) to the magnetic energy ($e_{\rm B}$) and the ratio of the true field strength to the equipartition value is
\begin{equation}
\frac{e_{\rm cr}}{e_{\rm B}} = \left( \frac{B_{\rm true}}{B_{\rm eq}} \right) ^{-\frac{p+5}{2}}, 
\label{eq:Btrue}
\end{equation}   
where $e_{\rm B}=B_{\rm true}^2/8\pi$. In our simulations, the CRe energy density needed to reproduce the observed haze emission is $e_{\rm CRe} \sim f_{\rm e,\nu} e_{\rm cr,sim} \sim (6\times 10^{-4})(1\times 10^{-10}) \sim 6\times 10^{-14}\ \mbox{ergs\ cm}^{-3}$. Assuming a CR proton-to-electron energy density ratio $k\equiv e_{\rm CRp}/e_{\rm CRe}=100$, we can use $e_{\rm cr}\sim 6\times 10^{-12}\ \mbox{ergs\ cm}^{-3}$ in Eq.\ \ref{eq:Btrue} and solve for the expected field strength. This gives $B_{\rm true} \sim 2.3\ \mu G$ (assuming $p=2$), consistent with the field strength at $z\sim 5-6$\ kpc (Figure \ref{fig:Bprf}). However, if the actual particle content is greater than the amount included in our simulations, then the true field strength could be smaller than the above estimate. These particles would need to be either hadrons or leptons of energies such that the corresponding synchrotron and IC emission would not contribute to the haze and gamma-ray bubble emission. 

The polarized lobe emission observed by {\it S-PASS} exhibits gradually {\it softer} (steeper) spectra with increasing $|b|$ \citep{Carretti13}, in contrast to the nearly uniform hard spectra of the {\it Fermi} bubbles and the {\it WMAP} haze. Moreover, the 2.3\ GHz lobes extend to $|b|\sim 60^\circ$, whereas both the {\it Fermi} bubbles and the {\it WMAP} haze present sharp edges at $|b|\sim 50^\circ$ \citep{Su10, Dobler12b}. This suggests the existence of an extra population of lower-energy CRs responsible for the polarized emission that is distinct from the CRs producing the {\it Fermi} bubbles and the {\it WMAP} haze \citep{Carretti13}. This unaccounted CR population implies a greater total particle energy, and thus suppressed magnetic fields compared to the simulated values. Moreover, we note that there appears to be a spatial anti-correlation between the polarized emission and the haze emission, which becomes more evident at higher $|b|$, and more so for the southern bubble.   

\begin{figure*}
\begin{center}
\includegraphics[scale=0.7]{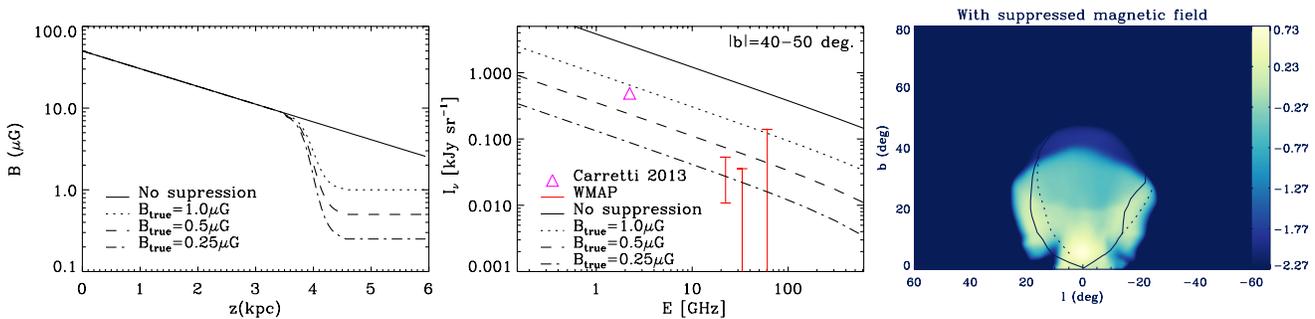} 
\caption{The deficient microwave emission at $|b|>35^\circ$ is possibly due to suppression of the magnetic fields, which is suggested by the existence of lower-energy CRs responsible for the polarized emission at $2.3$\ GHz. {\it Left:} Models of magnetic field distribution for different levels of suppression at $|z|>3.5$\ kpc. {\it Middle:} Microwave spectra for different levels of suppression of magnetic fields. {\it Right:} Map at 23\ GHz with magnetic field suppression with $B_{\rm true}=0.25\mu G$.} 
\label{fig:Bsup}
\end{center}
\end{figure*}

Figure \ref{fig:Bsup} illustrates how the properties of the microwave emission would change assuming different levels of magnetic field suppression. To mimic this effect, we generate the microwave map and spectra using artificial magnetic field distributions as shown in the left panel. These models assume field strength of $B_{\rm true}$ for $|z|>4.5$\ kpc, i.e., above the vertical distance of 4.5\ kpc away from the Galactic plane. A smooth transition (using the $\tanh$ function) is applied to connect the assumed $B_{\rm true}$ at $|z|=4.5$\ kpc and the exponential model field value at $|z|=3.5$\ kpc. Since the true distribution is unknown, these choices are arbitrary and are only meant to show possible effect of magnetic field suppression. 

As shown in the middle panel of Figure \ref{fig:Bsup}, the amplitude of the modeled microwave spectra at $|b|=40^\circ-50^\circ$ decreases significantly after including the suppression of the magnetic field. In particular, the model for $B_{\rm true}=0.25\ \mu G$ (dash dotted line) is in much better agreement with the deficient {\it WMAP} haze radiation than the model without magnetic field suppression (solid line). Compared to the map without suppression (top left panel of Figure \ref{fig:2jets}), the deficiency of the haze emission at $|b|>35^\circ$ can be easily seen in the simulated microwave map at 23\ GHz when the magnetic field suppression is included (right panel of Figure \ref{fig:Bsup}). The morphology of the simulated microwave haze including the magnetic field suppression is similar to that observed. We note that the gamma-ray map is not affected by the lower-energy CR population, because the bubble emission at high $|b|$ is dominated by CRe with energies higher than $\sim 100$\ GeV \citep{Dobler12a}, while the characteristic energy of the CRs causing the 2.3\ GHz polarized emission is $\sim 20$\ GeV \citep[][Eq.\ 2, assuming $B=0.25\ \mu G$]{Beck05}. 

According to Eq.\ \ref{eq:Btrue}, the field strength of $B_{\rm true}=0.25\ \mu G$ implies total particle energy density of $e_{\rm cr}\sim 1.7 \times 10^{-10}\ \mbox{ergs\ cm}^{-3}$ for $p=2$, much larger than the estimated particle energy density that generates the bubble and haze emission. These extra particles could thus be a possible source of the excess 2.3\ GHz polarized radiation in a model for $B_{\rm true}=0.25\ \mu G$ (middle panel of Figure \ref{fig:Bsup}). The actual level of the excess is difficult to estimate though, because the spectrum of the lower-energy particles is unknown. Future observations in the radio waveband will be very helpful in determining the spectrum of the lower-energy CR population and will enable detailed modeling of their observable signatures. 

The total particle energy density estimated above corresponds to a total particle pressure of $P_{\rm cr}=(\gamma_{\rm cr}-1)e_{\rm cr}\sim 6\times 10^{-11}\ \mbox{dyn\ cm}^{-2}$, where $\gamma_{\rm cr}=4/3$ is the effective adiabatic index of relativistic CRs. This is about half of the total pressure contained in our simulated bubbles, and hence could have a non-negligible dynamical effect and modify the magnetic field distribution. Therefore, a separate set of numerical simulations that self-consistently account for the dynamical evolution of the low-energy CRs is needed to further corroborate the possibility that the latitudinal variations in the magnetic field strength could account for the decrease of microwave emission at high $|b|$. 

\subsection{Compositions of the bubbles and the AGN jets}


In this paper we have assumed that all of the {\it Fermi} bubble and microwave haze emission originates from {\it leptonic} CRe contained in the AGN jets. However, alternatively the emission could also come from the gamma-ray photons and the synchrotron radiation from secondary electrons and positrons produced during the hadronic process involving CRp. Using the hadronic model (assuming the same energy range and slope for the CRp spectrum as for the default CRe spectrum), we find that roughly all of the CR energy density in our simulations is needed to match the observed gamma-ray spectrum, i.e., $f_{\rm p,\gamma}\sim 1$, in accordance with the previous result of \cite{Guo12a}. The detailed distribution of the secondary CRe requires following their generation and transport during the simulations, which will be a part of our future work to disentangle the relative importance of the leptonic and hadronic models. To the first order, we estimated the amount of secondary CRe by post-processing the CR energy densities and thermal gas densities in the simulation outputs using the method of \cite{Kelner06}. Although at any instant the production rate of the secondary CRe is small, we find that at the end of the simulations their {\it cumulative} number density becomes comparable to the amount of primary CRe estimated in the leptonic model. Therefore, for the purely leptonic scenario to be valid, $f_{\rm p,\gamma} \ll 1$ is required in order for the gamma-ray and microwave emission produced by the hadronic model to be subdominant. 

The above simple estimates have two interesting implications for the composition of the {\it Fermi} bubbles. Firstly, accordingly to our analyses of the leptonic model, the required amount of CRe with respect to the total simulated CR energy density is $f_{\rm e,\gamma} \sim 5 \times 10^{-4}$. Assuming the most conservative limit of $f_{\rm p,\gamma} \leqslant 0.1$, the leptonic model requires an upper limit for the CR proton-to-electron energy density ratio, i.e., $k \leqslant 200$, both for the AGN jets and inside the bubbles (since our simulations do not include other CR production and loss mechanisms). Having a limit of $k$ consistent with the range determined for radio bubbles in galaxy clusters \citep{Dunn05}, the {\it Fermi} bubbles indeed could be a Milky-Way analog of extragalactic radio bubbles/lobes inflated by AGN jets. Note also that given the same spectral ranges for both the CRe and CRp in our estimates, the CR proton-to-electron {\it number} density ratio should also be around $K \leqslant 200$. The typical value of $K \sim 100$, well-motivated by theories and CR measurements near the Sun \citep[][and references therein]{Beck05}, is consistent with the derived limit.  



Secondly, from constraints of the dynamical expansion and morphology of the bubbles, our previous simulations have yielded an estimate for the total pressure inside the bubbles, $P_{\rm tot}=P_{\rm th}+P_{\rm cr}\sim 10^{-10}\ \mbox{dyn\ cm}^{-2}$, in which the thermal pressure $P_{\rm th}$ and the total CR pressure $P_{\rm cr}$ (with no distinction between CRe or CRp) are degenerate (Y12; note that the magnetic pressure is subdominant except near the GC and in the filaments). This degeneracy could now be broken because we can independently constrain the contribution of CRe to $P_{\rm tot}$ using the gamma-ray and microwave emission. Our results show that the CR pressure of leptonic CRe needed to produce the bubble and haze emission is $P_{\rm CRe} = (\gamma_{\rm cr}-1)e_{\rm CRe} \sim 2 \times 10^{-14}\ \mbox{dyn\ cm}^{-2}$, and therefore the total particle pressure is $P_{\rm cr}=(1+k)P_{\rm CRe} \leqslant 2 \times 10^{-12}\ \mbox{dyn\ cm}^{-2}$, which is $\leqslant 2\ \%$ of the total pressure inside the bubbles. This implies that in the {\it purely leptonic} scenario, most of the pressure within the {\it Fermi} bubbles comes from the thermal pressure. For cluster radio bubbles, the existence of X-ray cavities and shocks usually argues against the bubbles being dominated by the thermal pressure \citep{McNamara07}. However, we note that even though a non-negligible amount of thermal gas exists within the {\it Fermi} bubbles, prominent limb-brightened features can still be seen in the 1.5\ keV X-ray map (Y12, Figure 6) because of the large gas compression factor of strong shocks (Mach number $M>10$), as opposed to the weak shocks associated with the buoyantly rising cluster bubbles. Moreover, while gas entrainment within AGN jets is usually neglected when deriving properties of cluster radio bubbles, the entrainment of gas (a jet density contrast of $\eta \sim 0.05$) is required to explain the elongated morphology of the {\it Fermi} bubbles (\cite{Guo12a}; Y12).  


\section{Conclusions}
\label{sec:conclusion}

The physical origin of the {\it Fermi} bubbles and the microwave haze is still under debate. The spatially resolved, broad-band properties of the {\it Fermi} bubbles revealed by the ample observational data from {\it Fermi} \citep{Su10, Hooper13}, {\it WMAP} \citep{Finkbeiner04, Dobler08}, {\it Planck} \citep{PlanckHaze}, and {\it S-PASS} \citep{Carretti13} have offered an unprecedented opportunity to disentangle different mechanisms. 

We investigate the properties of the {\it Fermi} bubbles and the microwave haze using 3D MHD simulations of CR injection from the central supermassive black hole assuming the leptonic model. Our previous simulations have successfully reproduced the primary features of the {\it Fermi} bubbles and the {\it ROSAT} X-ray arc features (Y12). In this work, we build upon the previous study by employing more realistic models for the magnetic field and by generating gamma-ray and microwave maps and spectra for direct comparisons with observations. We focus on identifying the critical physical mechanisms that are responsible for the {\it spatial, spectral, and polarization} properties of the bubble and haze emission. Our findings are as follows.

1.\ The same population of leptonic CRs can simultaneously account for the bubble and haze emission provided that the magnetic fields within the bubbles are close to the exponentially distributed Galactic magnetic field. This can be explained by mixing from the ambient field followed by turbulent field amplification.

2.\ Since the required field strength within the bubbles is sufficient to push the gas and CRs away from the GC, the observed centrally peaked profile of the {\it WMAP} haze suggests replenishment of CRs at low $|b|$. If the new {\it Fermi} data confirms the pair of gamma jets recently discovered by \cite{Su12}, such jets could supply the CRs needed to explain the rising microwave profile toward the GC.    

3.\ A high degree of polarization is expected in the bubble interior due to ordered magnetic fields stretched radially by elongated vortices behind the shocks. For the bubble exterior, the regularly structured fields aligned with the bubble surface by magnetic draping could also provide the necessary condition for highly polarized signals if other sources of CRs are present outside the bubbles. These simulation predictions are broadly consistent with the polarization measurements at 2.3\ GHz \citep{Carretti13}. 

4.\ Enhancement of rotation measures between the bubble edges and the shock fronts is expected because of increased gas density as well as more amplified and ordered magnetic fields. The exact location and level of enhancement, though, depend on projections and the actual field geometry.  

5.\ The properties of the 2.3\ GHz polarized lobe emission suggests the existence of lower-energy CR population that is distinct from the CRs producing the bubbles and the haze. We show that the existence of these extra CRs is consistent with magnetic field suppressed below equipartition. This magnetic field suppression likely causes the decrease in the microwave haze emission at $b<-35^\circ$.   

6.\ In the purely leptonic scenario (in which contribution from CRp to the radiation is subdominant), we obtained a limit on the CR proton-to-electron energy density $k \leqslant 200$. Combined with the dynamical constraints on the total bubble pressure from the simulations, we find that the pressure within the bubbles is dominated by the thermal gas in the purely leptonic model.


Our simulations currently do not include possible mechanisms of CR acceleration and energy losses due to IC and synchrotron radiation and adiabatic expansion. As discussed in Y12, due to the effect of adiabatic expansion, the CR spectrum at the early stage of the bubble evolution must have higher energies than observed today. The higher CR energies and stronger ISRF and magnetic fields near the GC when the bubbles were young imply significantly shorter CR cooling time and steepening or curvature in the CR spectrum. The fact that the observed hard gamma-ray spectrum of the {\it Fermi} bubbles does not show significant evidence of cooling indicates that the cooling losses may have been offset by CR acceleration close to the origin of the jets. Our future study will incorporate these processes using detailed modeling of the CR spectrum, which will provide valuable information on the required CR source spectrum and acceleration processes during the bubble evolution.    


\section*{Acknowledgments}

The authors would like to thank Douglas Finkbeiner, Greg Dobler, Meng Su, William Mathews, Fulai Guo, and Peng Oh for helpful discussions. We thank Hui Li for helping to assess the importance of the hadronic model. HYKY and MR acknowledge the NSF grant AST 1008454. MR acknowledges NASA ATP grant (12-ATP12-0017). EZ acknowledges support from NSF grants PHY 0821899 and AST0903900. The simulations in this study are carried out using the resources of the Texas Advanced Computing Center (allocation TG-AST120065). FLASH was developed in part by the DOE NNSA ASC- and DOE Office of Science ASCR-supported Flash Center for Computational Science at the University of Chicago.


\bibliography{fermi}



\appendix

\section{Gamma-ray emission from mixed-in cosmic ray protons}
\label{appendix2}

As discussed in \S~\ref{sec:Bfield}, the magnetic field inside the bubbles is intially mixed in from the ambient field and then amplified by turbulent eddies. Such mixing could potentially bring the interstellar CRp (ISM CRp) into the bubbles and produce gamma-ray emission via the hadronic process. In this appendix we estimate the contribution to the gamma-ray emission by these mixed-in CRp in order to see whether they could contaminate the bubble emission produced by the CRe contained in the AGN jets. This can be estimated by calculating the relative hadronic emission from the mixed-in CRp to the ISM CRp, because the latter is comparable to the observed bubble emission (see Figure 12 in \cite{Su10}).  

For hadronic emission, the gamma-ray emissivity is directly proportional to the CR number density ($n_{\rm cr}$) and the gas number density ($n_{\rm e}$). Therefore, the ratio between the gamma-ray surface brightness ($I$) of the mixed-in CRp and the ISM CRp can be estimated as the following:

\begin{equation}
\frac{I_{\rm mixed\ CRp}}{I_{\rm ISM\ CRp}} = \frac{\int_{r<r_{\rm bub}} n_{\rm cr,mix}(r)n_{\rm e}(r) dl}{\int n_{\rm cr,ISM}(r)n_{\rm e}(r) dl}, 
\end{equation}
where $r_{\rm bub}$ is the radius of the bubbles. Choosing the CR number density at the bubbles edges (e.g., at the top) as a reference point and defining $f_{\rm mix}(r) = n_{\rm cr,mix}(r)/n_{\rm cr,ISM}(R=0,z=6)$ and $f_{ncr}(r) = n_{\rm cr,ISM}(r) / n_{\rm cr,ISM} (R=0,z=6)$, where $R$ is the distance to the rotational axis of the Galaxy, then

\begin{equation}
\frac{I_{\rm mixed\ CRp}}{I_{\rm ISM\ CRp}} = \frac{\int_{r<r_{\rm bub}} f_{\rm mix}(r)n_{\rm e}(r) dl}{\int f_{\rm ncr}(r)n_{\rm e}(r) dl}, 
\end{equation}
where the mixing fraction $f_{\rm mix}(r)$ is measured from our test runs with tracer particles, and the normalized distribution of ISM CRp, $f_{\rm ncr}(r)$, can be computed assuming $n_{\rm cr,sim}(r) \propto B(r)^2$ and $B(r)$ follows the exponential model. 

The ratio we obtained by performing the above integration is $I_{\rm mixed\ CRp} / I_{\rm ISM\ CRp}=1.5\times10^{-2}$. The ratio is small because (a) $f_{\rm mix} < 1$ while $f_{\rm ncr}$ can be $> 1$ near the Galactic plane; (b) Mixing is non-negligible ($f_{\rm mix}>0.01$) only for a very small path length, $d\sim 1-2$\ kpc, while $f_{\rm ncr}>0.01$ for a much longer path length $D\sim14$\ kpc. Therefore, we conclude that the contamination by the mixed-in ISM CRp is small compared to the bubble emission.

\section{Convergence Study}
\label{appendix}

In \S~\ref{sec:Bfield}, we showed that for the leptonic CRs to simultaneously reproduce the {\it Fermi} bubble and microwave haze emission, the magnetic field inside the bubbles has to be comparable to the ambient exponential distribution. This can be achieved if the magnetic field inside the bubbles is first partially mixed in from the ambient medium, and is subsequently amplified by turbulent eddies in the post-shock region. In this case, the final bubble field strength is mainly determined by the process of amplification instead of the level of mixing. 

In order to examine the robustness of the results, we performed additional simulations to see whether the simulated magnetic field inside the bubbles is convergent when we increase the resolution. We focus on the convergence of the $l_{\rm B}=1$\ kpc simulations because it is the more physically relevant case for producing the large bubble field needed to explain the microwave haze emission. For the convergence tests, we use a constant initial field with an average strength of $1\ \mu G$ instead of the exponential field because the large field strength in the latter case is more computational difficult for simulations with higher resolution. We turned off the magnetic field injection in order to focus on the evolution of the bubble field due to turbulent amplification. We use the HLL Riemann solver for the results presented in this paper, but we did verify that the results are not sensitive to the choice (e.g., HLLD or HLL). Finally, the initial gas density profile is softened within a radius of the jet width (0.5\ kpc) so that the same jet parameters can be used for the higher-resolution simulations without altering the shape of the bubbles.  

\begin{figure}
\begin{center}
\includegraphics[scale=0.85]{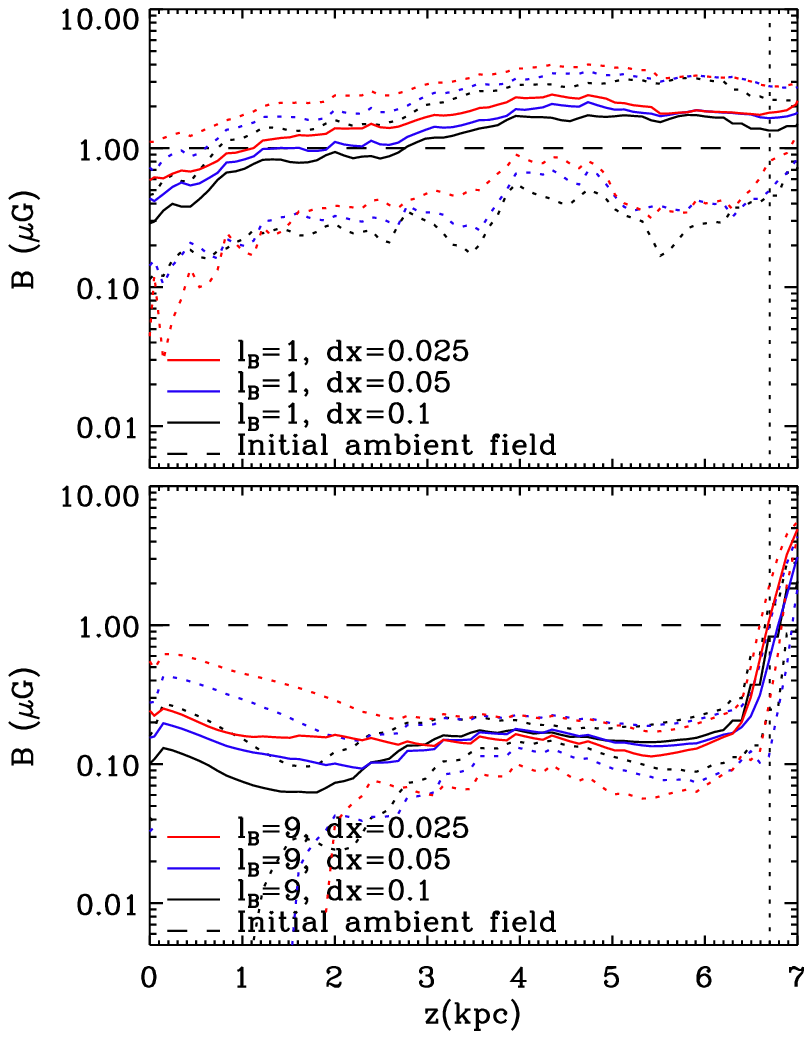} 
\caption{Simulated field strength at $t=1.2$\ Myr as a function of vertical height from the Galactic plane for simulations with different resolution ($dx$ is the size of the finest grid cell in units of kpc). The top and bottom panels show results using initial magnetic field coherence lengths $l_{\rm B}=1$\ kpc and $l_{\rm B}=9$\ kpc, respectively. The solid lines represent the field strengths averaged within the region $|x|<1$\ kpc and $|y|<1$\ kpc, and the dotted lines are the corresponding standard deviations. The vertical dotted line shows the height of the bubbles at $t=1.2$\ Myr (see the text for definition). } 
\label{fig:resol}
\end{center}
\end{figure}

\begin{figure}
\begin{center}
\includegraphics[scale=0.85]{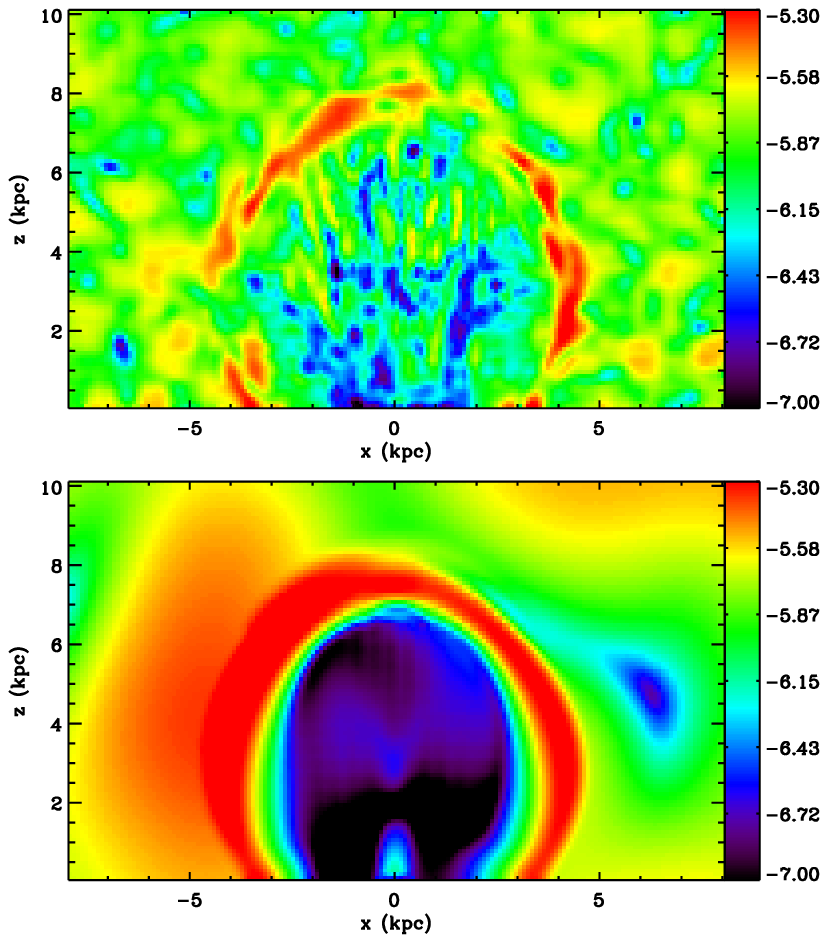} 
\caption{Maps of magnetic field strength (plotted in logarithmic scale in units of Gauss) for runs with initial magnetic field coherence length $l_{\rm B}=1$\ kpc (top) and $l_{\rm B}=9$\ kpc (bottom) at $t=1.2$\ Myr. The grid resolution of these simulations is 0.1\ kpc.} 
\label{fig:Bmap}
\end{center}
\end{figure}

Figure \ref{fig:resol} plots the magnetic field strength at $t=1.2$\ Myr as a function of the vertical height from the Galactic plane for different resolutions and initial magnetic field coherence length $l_{\rm B}$. The solid and dotted lines show the average and standard deviation, respectively, of the field strengths inside the region $|x|<1$\ kpc and $|y|<1$\ kpc. 
The results of the convergence tests suggest that the magnetic field inside the bubbles is convergent in some regions (high $|z|$) and close to convergence in others (low $|z|$). The fact that even for the latter case the results differ by very little gives us confidence that the results are physically meaningful and that our main conclusions remain independent of the resolution.
The vertical dotted line shows the height of the bubbles, $z_{\rm bub}$, which is defined to be the maximum height for which the CR energy density of all the grid cells within the selected region is greater than a chosen threshold, i.e., $e_{\rm cr} \geqslant 1\times 10^{-12}\ \mbox{ergs\ cm}^{-3}$. Note that the height of the bubbles in the resolution tests ($z_{\rm bub}\sim 6.7$\ kpc) is greater than the bubble height obtained for the simulations presented in the main text ($z_{\rm bub}\sim 6.0$\ kpc) because of the different initial gas density profiles adopted.
Figure \ref{fig:Bmap} shows the magnetic field distribution at the end of the lower-resolution simulations for different $l_{\rm B}$.
The difference between simulations with varying $l_{\rm B}$ can be clearly seen from both figures regardless of the resolution: the magnetic field within the bubbles is weaker than the ambient field for $l_{\rm B}=9$\ kpc, whereas the bubble field strength is comparable to the ambient value for $l_{\rm B}=1$\ kpc, in accordance with the trend found in \S~\ref{sec:Bfield}.     

As discussed in \S~\ref{sec:Bfield}, the difference in the bubble field strength for different $l_{\rm B}$ results primarily from the timescale for turbulent amplification, $t_{\rm eddy}=l/\sigma_{\rm sh}$. More specifically, $\sigma_{\rm amb}$ measured on a scale of $l=0.1$\ kpc is $\sim 30\ \mbox{km\ s}^{-1}$ for $l_{\rm B}=1$\ kpc and $\sim 1.5\ \mbox{km\ s}^{-1}$ for $l_{\rm B}=9$\ kpc. Assuming $\sigma_{\rm sh}=\sqrt{2} \sigma_{\rm amb}$ \citep{Larsson09}, the corresponding eddy turnover time is $t_{\rm eddy} \sim 2$\ Myr for $l_{\rm B}=1$\ kpc and twenty times longer for $l_{\rm B}=9$\ kpc. The growth of the bubble field can be described as $B_{\rm f}=B_{\rm i} \exp (t/t_{\rm eddy})$, where $B_{\rm i}$ and $B_{\rm f}$ are the field strengths before and after turbulent amplification, respectively. Within the bubble formation time $t\sim 1.2$\ Myr, the magnetic field can grow by a factor of $\sim 2$ for $l_{\rm B}=1$\ kpc and only $\sim 1.05$ for $l_{\rm B}=9$\ kpc. Note that the difference in the final bubble field strength in these two cases is greater than what is implied by the amplification factors. This is because the effect of amplification is cumulative, i.e., once the bubble field is mixed in and amplified at earlier stages of the bubble evolution, later amplification becomes easier because $B_{\rm i}$ is already large. Another thing to note is that, for $l_{\rm B}=1$\ kpc, the fact that the final field strength is comparable to $\sqrt{8}$ of the ambient value (see Eq.\ \ref{eq:equip}) implies that the bubble field has nearly saturated, while the field is still being amplified at the end of the simulations for $l_{\rm B}=9$\ kpc. 

\begin{figure}
\begin{center}
\includegraphics[scale=0.85]{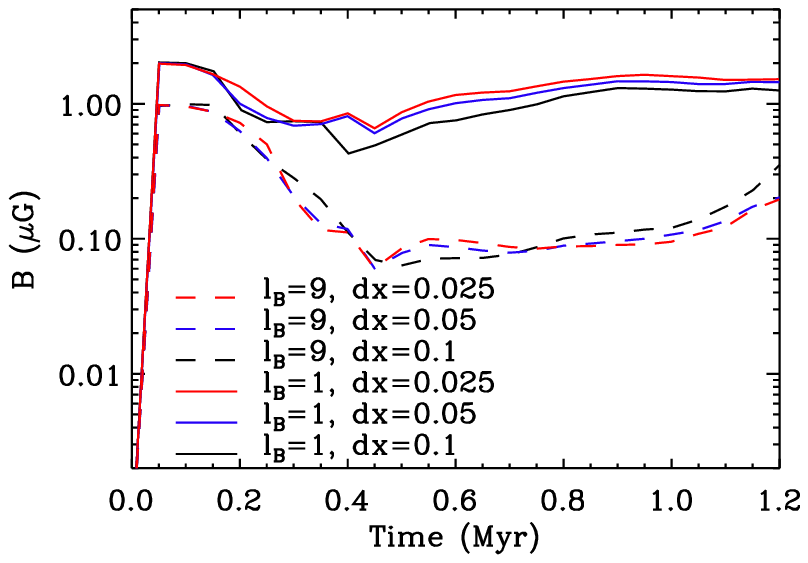} 
\caption{Evolution of magnetic field strength near the bubble edges for different resolution $dx$ and initial magnetic field coherence length $l_{\rm B}$. The field strength is averaged within the region $|x|<1$\ kpc, $|y|<1$\ kpc, and $z_{\rm bub}-1<|z|<z_{\rm bub}$, where $z_{\rm bub}$ is the height of the bubbles at a given time in units of kpc (see the text for definition).} 
\label{fig:Bevol}
\end{center}
\end{figure}

These processes are illustrated in Figure \ref{fig:Bevol}, which shows the evolution of magnetic field near the bubble edges. The magnetic field starts from the ambient field value that were mixed in (as well as amplified for $l_{\rm B}=1$\ kpc) before the first output file at $t=0.05$\ Myr. It then decreases as the bubbles expand adiabatically at $t\sim 0.2-0.4$\ Myr, and is amplified again until the end of the simulations. Because the amplification timescale is shorter for $l_{\rm B}=1$\ kpc than for $l_{\rm B}=9$\ kpc, the field strength saturates to the value determined by energy conservation for $l_{\rm B}
=1$\ kpc, while the field is still being amplified at the end of the simulations for $l_{\rm B}=9$\ kpc.

The results for different resolutions in general show good convergence (i.e., the results for $dx=0.025$\ kpc and $dx=0.05$\ kpc are closer to each other than those for $dx=0.05$\ kpc and $dx=0.1$\ kpc) for both the final field strength within the bubbles (Figure \ref{fig:resol}) and the evolution of magnetic field (Figure \ref{fig:Bevol}). \footnote{There is an implicit dependence on numerical resolution, however, to the extent that the degree to which fields are mixed into the bubble in the first place depends on diffusion. If the actual mixing were smaller, the amplification time would be longer. Since turbulent mixing and magnetic reconnection are both difficult to quantify, we assume that the mixing is efficient enough to provide the seed for amplification as observed here.} For $l_{\rm B}=1$\ kpc, the field strength near the bubble edges is convergent because the magnetic field has already saturated; increasing the resolution only shortens the timescale for amplification (since $t_{\rm eddy} \propto l^{2/3}$ for Kolmogorov turbulence) but does not affect the final field strength. For regions further away from the shocks (i.e., for lower $|z|$), the field strength slightly increases with resolution, which is consistent with the expectation that the post-shock vortices dissipate on a shorter timescale for the lower-resolution simulations. We note that the slightly stronger bubble field in the higher-resolution simulations strengthens the need for CR replenishment by the second jet.

\label{lastpage}

\end{document}